# Understanding the leakage process for multi-scale water infrastructure asset management: necessity for a dialogue between sociological and data sciences

Marie Collet*^, Nicolas Rodriguez*, Selma Baati*, Alain Husson*, Eddy Renaud*, Kévin Caillaud* & Yves Le Gat*

^Corresponding author (marie.collet@inrae.fr)

*INRAE Nouvelle-Aquitaine Bordeaux, UR ETBX, Cestas Gazinet, France

## Abstract

Reducing water losses is one of the most pressing issues for modern water utilities. To that end, improving the efficiency of the pipe leakage and repair process and aiding the selection of the pipes that are to be renewed or rehabilitated are essential. To help addressing these tasks, in this work, we develop a model predicting the probability of a pipe to be leaking. This work is set the context of a multidisciplinary project with Société Wallone des Eaux and it is aligned with their goal to improve their Infrastructure Asset Management in the short and the long terms. Developing and feeding this leakage probability model relies on an intense data processing phase, mobilizing data and water engineering sciences, since the raw data from SWDE is not ready to be used in the model. Complementarily, we thus employ techniques from sociology (e.g., interviews, analyses of the human/non-human actors and of the tools, sociotechnical translations) in order to complete the data, to improve our understanding of its production, and to increase its value and its availability for the prediction of the pipe leakage probability. This model will be implemented in SWDE's information system and used for strategies to reduce water losses.

## Keywords

## Introduction

Water supply systems are designed to provide drinkable water with a flow and pressure that should be sufficient for each point of consumption. Water losses occur mainly due to pipe leaks, which are caused by deteriorating infrastructures. They may affect the water distribution network performance and therefore, may lead to an increase in the abstraction of raw water from drinkable water sources. One of the most pressing issues in the management of water utilities is to reduce leaks in order to ensure a sustainable use of water resources (Hafskjold, 2010; Renaud and Charriere, 2016). Thus, water systems operators are moving towards developing strategies to control leakages in pipe networks. Apart from service pressure reduction or modulation, such strategies are focused on two major types of action: the implementation of research campaigns which aim at leaks detection and repair as soon as possible; and rehabilitation/renewal of parts of the distribution network which are characterized by the highest frequencies of leakage occurrence (Renaud et al., 2014, 2012; Rokstad, 2012). The cost of those shares, especially the expense of replacing pipes, may raise the operating cost of water utility, hence water price. Thus, the implementation of an efficient leakage control strategy could be based on tools that assist in the selection of actions that are better suited for water utility management and its goals for short and long terms. An example could be a leak prediction model for water distribution networks, for which an efficient and well-documented water leakage (or failure) detection and repair process generating accurate data is essential.

Data from many sources and in varying formats are collected, stored, and used by water utilities for most of their administrative and business processes (e.g., data on the hydraulic characteristics of the network, its structure, data on the network environment, failure history). Infrastructure Asset Management (IAM) relies on detailed and structured technical information on the water infrastructures and operations involved in supplying water to users, but even more on the understanding of the rationale behind the organization of water utilities and the work practice of its many employees. Yet, these data, when available, are not tailored for the construction of probabilistic models. This is often due to their production and exploitation methods (processing to which they are subject, archiving methods, recording in databases, etc.). Therefore, the pipe leakage state may not be obvious in the raw data, leading to uncertainties.

Based on the experience from an applied research project, called GePaME (Multi-scale IAM), led in collaboration between Société Wallonne des Eaux (SWDE) and INRAE-Bordeaux, this paper presents a set-up of research tasks designed to extract pipe maintenance (e.g., inspection, repair) data embedded in SAP (Enterprise Resource Planning software) records and Geographical Information Systems (GIS) data from SWDE in order to infer a pipe leakage probability model. The goal of this model is to help SWDE improve its long-term IAM process. The originality of this work lies particularly in the link between pipe leakage detection and repairs, as well as in the inter-disciplinary work involved to process the available raw data.

**Methods**

In fact, the raw data provided by SWDE cannot be used to supply a probabilistic model. In order to make better use of these data, we propose to study and analyse its production and exploitation process in it whole sociotechnical dimension, meaning: actors, tools and technical artefacts (software, databases, instruments, etc.), and current practices in leaks detection and repair, in order to complete and transform existing raw data into information suitable for IAM. This analysis is be partly based on planned interviews with actors involved in generating the raw data at different levels.

For this, analyses mobilizing four disciplines are combined: (i) Data science, to cleanse and connect network segment description and localisation data to leakage detection operations and pipe leak repairs; (ii) Drinking water engineering science, to exploit DMA monitoring data; (iii) Sociology (sociology of Sciences and Technology, sociology of work) to understand how DMA, and pipes within a DMA, are practically selected for leakage inspection, and how inspection data are produced and reported; (iv) Probability and statistics, for building a modelling tool that could account for this cumulated knowledge and the associated uncertainties.

Mainly three problems can be a hindrance to exploiting the vast amount of data that water utilities generate on a daily basis for IAM (Okwori et al., 2021; Rokstad, 2012). First, water utilities may not always store the crucial metadata describing how, when, and by whom the data were recorded (or updated) and validated. Therefore, the data can contain redundancies and inconsistencies, making the accuracy and priority given to the information not always clear. In addition, various technical terms (e.g., District Metered Area (DMA), pipe section, leak, and failure) can have different meanings for different operators involved in the data production. Second, many technical details linked with work practice and experience are not reported due to the administrative procedures that historically framed the data collection (data overwritten or not archived). Again, extracting information useful for IAM from raw data is difficult, even with the most advanced data mining techniques. Finally, the data may not be structured as a unique and consistent database (Rokstad, 2012). Ideally, all technical objects (such as pipes) and technical tasks (such as pipe leak repairs) are represented as entities, that is, tables with primary keys (IDs which are not duplicated nor null), attributes (characteristics) which are categorized (not free text), and foreign keys (links to other entities). This structure allows a seamless use of the data by merging different tables to create information useful for probabilistic modelling.

We therefore develop advanced data processing algorithms that overcome these difficulties inherent in the data, with the R software language (R Core Team, 2020).

We eventually develop a statistical model of the pipe leakage state probability based on the Linear Extension of the Yule Model with selective survival (LEYP, Le Gat, 2014) and feed it with the processed data combining the network characteristics, DMA, and the leakage detection and pipe leak repair (technical and sociological data).

**Results**

This multidisciplinary approach allows a better understanding of the technical and social context in which SWDE's leak detection strategy is employed. The production and exploitation of the data are indeed oriented by the finalities of leak inspection process (regulatory, environmental, economic, administrative, etc.). This process implies many tools/instruments used to generate data and individuals/organizations to implement these tools and interpret these data. Taken into consideration, the production and exploitation data can be presented as a "chain of translations" which links human and non-human entities (Akrich et al., 2006). By moving from one entity to another during production and operation, data undergoes transformations and takes on meaning. Actually, these issues of finalities and the attribution meaning are not known very well.

From the clarification of the chain of translations related to leakage and break detection and repair, we expect a better understanding of how to filter, cleanse and interconnect available raw data in order to identify which DMA, and which pipes inside these DMA's, were inspected and when, and which pipes among these were found to be subject to leakage or breakage. This will then make it possible to build a leakage probability model, as well as a break intensity model, both usable within a long term simulation framework devoted to assess the relevance of pipe inspection and renewal strategies.

This mobilization of inter-disciplinary skills requires a common vocabulary and thus sharing different epistemological cultures, from the field agents to the data scientists, via sociology of work. The study of the translation chains along the data course adds essential information that was not recorded in the current database. Assessing the adequacy of this database design and identifying its areas of improvement on a short and long-term perspective for IAM heavily depends on a clear understanding of the organization of SWDE on both on a governance level and on an operational level, which pertains again to sociological practice.

Finally, the shift of temporalities (from daily basis for operational work on the pipe to long-term strategies of network renewal) is taken into account to make sure that the SWDE will be able to make use of and adapt the tools produced and integrate them in their daily practice of data acquisition.

**Conclusion**

This study thus highlights the need to better understand the processes that underlie the leakage detection operations both technically and sociologically. On the one hand, the leakage detection operations could answer multiple objectives (administrative, financial, technical, regulatory…) that may influence the technics used and/or the data stored. On the other hand, these detection operations are carried out by human field operators, and the question of the impact of their perception and understanding of where and how to implement an operation remains open. The dialogue between sociology, data science, water supply engineering and probability is also at the core of the comprehension of the mutual influence between the technical tools and the realization of the leakage detection operations, as well as their evolutions in order to take into account possible biases in the statistical model.